\documentclass{pasa}

\usepackage{rotating, graphicx}
\usepackage{multirow}
\usepackage{caption, subcaption}
\usepackage{pdflscape}

\usepackage{lmodern}

\usepackage[T1]{fontenc}

\usepackage{textcomp}

\newcommand{\noopsort}[1]{}
\def\h {^{\mathrm{h}}}
\def\m {^{\mathrm{m}}}
\def\deg {^{\circ} }
\def\sqdeg {\,deg$^2$}
\def\ujybm {\,$\mu$Jy/beam }

\def\_ {\textunderscore}

\title[GLEAM South Galactic Pole data release]{GaLactic and Extragalactic All-sky Murchison Widefield Array (GLEAM) survey III: South Galactic Pole data release}

\author[Franzen et al.]{T.~M.~O.~Franzen$^{1,2}$\thanks{Email: franzen@astron.nl} ,
N.~Hurley-Walker$^{1}$,
S.~V.~White$^{1,3}$,
P.~J.~Hancock$^{1}$,
N.~Seymour$^{1}$,
A.~D.~Kapi\'{n}ska$^{4,5}$,
L.~Staveley-Smith$^{4,6}$
and R.~B.~Wayth$^{1,6}$
\affil{$^1$International Centre for Radio Astronomy Research, Curtin University, Bentley, WA 6102, Australia}
\affil{$^2$ASTRON: the Netherlands Institute for Radio Astronomy, PO Box 2, 7990 AA, Dwingeloo, The Netherlands}
\affil{$^3$Department of Physics and Electronics, Rhodes University, PO Box 94, Grahamstown, 6140, South Africa}
\affil{$^4$International Centre for Radio Astronomy Research, University of Western Australia, Crawley 6009, Australia}
\affil{$^5$National Radio Astronomy Observatory, 1003 Lopezville Rd, Socorro, NM 87801, USA}
\affil{$^6$ARC Centre of Excellence for All Sky Astrophysics in 3 Dimensions (ASTRO 3D), Australia}
}

\jid{PASA}
\doi{10.1017/pas.\the\year.xxx}
\jyear{\the\year}

\usepackage{aas_macros}
\usepackage{hyperref} 
\hypersetup{colorlinks,citecolor=blue,linkcolor=blue,urlcolor=blue}

\hypersetup{draft}

\begin{document}

\begin{frontmatter}
\maketitle

\begin{abstract}
We present the South Galactic Pole (SGP) data release from the GaLactic and Extragalactic All-sky Murchison Widefield Array (GLEAM) survey. These data combine both years of GLEAM observations at 72--231~MHz conducted with the Murchison Widefield Array (MWA) and cover an area of 5,113$~\mathrm{deg}^{2}$ centred on the SGP at $20\h 40\m < \mathrm{RA} < 05\h 04\m$ and $-48\deg < \mathrm{Dec} < -2\deg$. At 216~MHz, the typical rms noise is $\approx 5$~mJy/beam and the angular resolution $\approx 2$~arcmin. The source catalogue contains a total of 108,851 components above $5\sigma$, of which 77 per cent have measured spectral indices between 72 and 231~MHz. Improvements to the data reduction in this release include the use of the GLEAM Extragalactic catalogue as a sky model to calibrate the data, a more efficient and automated algorithm to deconvolve the snapshot images, and a more accurate primary beam model to correct the flux scale. This data release enables more sensitive large-scale studies of extragalactic source populations as well as spectral variability studies on a one-year timescale.
\end{abstract}

\end{frontmatter}


\section{Introduction}\label{Introduction}

Over the past decade, a number of new low-frequency ($\lesssim 200$~MHz) radio telescopes have come online, including the Low Frequency Array \citep[LOFAR;][]{vanhaarlem2013}, the Precision Array for Probing the Epoch of Reionisation \citep[PAPER;][]{parsons2010}, the Long Wavelength Array \citep[LWA;][]{ellingson2013} and the Murchison Widefield Array \citep[MWA;][]{tingay2013}, as part of preparations for the low-frequency Square Kilometre Array (SKA1 LOW). The design of these telescopes was largely guided by the goal of measuring the redshifted 21~cm signal from the Epoch of Reionisation (EoR), predicted to lie between 50 and 200~MHz \citep[e.g.][]{furlanetto2006}. Other science objectives of these telescopes include exploring the transient radio sky, performing Galactic and extragalactic surveys, and tracking solar, heliospheric and ionospheric phenomena.

The large-area sky surveys performed by these instruments allow the statistical properties of large samples of radio galaxies to be investigated, thereby contributing to our understanding of active galactic nucleus (AGN) and star formation activity over cosmic time \citep[e.g.][]{simpson2017}. Galaxies across the Universe, including our own Milky Way, produce strong foreground contamination in experiments seeking to detect the EoR. Accurate characterisation and removal of these foreground contaminants is a critical step in the interpretation of EoR data \citep[e.g.][]{procopio2017,trott2019}.

The GaLactic and Extragalactic All-sky Murchison Widefield Array \citep[GLEAM;][]{wayth2015} survey is an all-sky continuum survey at 72--231~MHz with an angular resolution of $\approx 2$~arcmin conducted with the MWA. The MWA is an interferometer operating at frequencies between 72 and 300~MHz, with an instantaneous bandwidth of 30.72~MHz. It is located at the Murchison Radio-astronomy Observatory in Western Australia and is a low-frequency precursor telescope for the SKA. The wide range of science enabled by GLEAM, much of which is dependent on the survey's wide areal coverage, large fractional bandwidth and high surface brightness sensitivity, is detailed in \cite{beardsley2019}.

The GLEAM Extragalactic data release \citep[Exgal;][]{hurleywalker2017} is based on the first year (2013--2014) of GLEAM observations. It covers the entire sky south of declination $+30\deg$ excluding the strip at Galactic latitude $|b| < 10\deg$, and a few regions affected by poor ionospheric conditions and around bright, complex sources, such as the Magellanic Clouds. The GLEAM Exgal catalogue contains 307,455 component sources above a 5$\sigma$ detection limit of $\approx 50~\mathrm{mJy/beam}$, the vast majority of which have measured in-band spectral indices. The sensitivity in GLEAM Exgal is limited by sidelobe confusion, i.e. noise introduced into the image due to the combined sidelobes of undeconvolved sources \citep{franzen2019}, while the flux density calibration is limited by errors in the primary beam model of order 5--20 per cent.

The GLEAM Exgal catalogue has been combined with higher-frequency radio catalogues to measure broad-band spectral energy distributions, enabling detailed studies of radio-loud AGN \citep{herzog2016,callingham2017} and local powerful star-forming galaxies \citep{kapinska2017,galvin2018}. The extremely high surface brightness sensitivity of GLEAM Exgal has made possible the detection and characterisation of large, faint objects such as dying radio galaxies \citep{duchesne2019}, and radio haloes and relics associated with galaxy clusters \citep{schellenberger2017}. \cite{franzen2019} used the GLEAM Exgal catalogue to derive low-frequency source counts above $\sim 100$~mJy to high precision, allowing tight constraints on bright radio source population models. More recently, \cite{white2020b,white2020a} constructed a complete sample of the `brightest' radio sources ($S_{151\,\mathrm{MHz}} > 4$~Jy) south of declination $+30\deg$ from the GLEAM Exgal catalogue, the majority of which are AGN with powerful radio jets; the GLEAM 4-Jy Sample is an order of magnitude larger than the 3CRR sample by \cite{laing1983} and will be a benchmark for the bright radio galaxy population.

In this paper, we present an extension to GLEAM in order to reach a sensitivity of $\approx 5~\mathrm{mJy/beam}$ in a $\approx$ 5,100~$\mathrm{deg}^{2}$ area of sky centred on the South Galactic Pole (SGP) at $20\h 40\m < \mathrm{RA} < 05\h 04\m$ and $-48\deg < \mathrm{Dec} < -2\deg$. The GLEAM SGP data release is based on a subset of the data from both years (2013--2015) of GLEAM observations. The region of sky covered in this release has been the target of a number of deep multi-wavelength surveys, such as the Galaxy and Mass Assembly \citep[GAMA;][]{driver2009} 02 and 23 fields, the \textit{Chandra} Deep Field South \citep[CDFS;][]{giacconi2001} and the European Large Area ISO Survey - South 1 \citep[ELAIS-S1;][]{oliver2000}. MWA observations dedicated to detect the first global signals from the EoR concentrate on two fields, named EoR0 (centred at $00\h -27\deg$) and EoR1 (centred at $04\h -30\deg$), which also lie in this region of sky \citep{beardsley2016}.

We analyse the GLEAM SGP data using an improved data reduction process which addresses some of the limitations of GLEAM Exgal, and present the GLEAM SGP images and source catalogue. The analysis of both years of GLEAM observations not only improves the image sensitivity, but also provides a second epoch which can be used to study spectral variability on a one-year timescale.

The layout of the paper is as follows. Section~\ref{GLEAM year 1 and 2 observations} outlines the survey strategy employed in GLEAM observations. Section~\ref{Data reduction} describes the improvements made to the GLEAM calibration and imaging pipeline in this release. Section~\ref{Source finding and cataloguing} outlines the steps taken to construct and validate the GLEAM SGP catalogue. Our results are summarised in Section~\ref{Summary}.

Throughout this paper, we assume the convention for spectral index, $\alpha$, where $S \propto \nu^{\alpha}$. Right ascension is abbreviated as RA and declination is abbreviated as Dec.

\section{GLEAM year 1 and 2 observations}\label{GLEAM year 1 and 2 observations}

The GLEAM year 1 and 2 observations were conducted with Phase I of the MWA. Phase I of the MWA consisted of 128 16-crossed-pair-dipole tiles with baselines extending to $\approx 3$~km. In this array configuration, using a uniform weighting scheme, the angular resolution at 154~MHz was $\approx 2.5$ by 2.2~sec($\delta + 26.7\deg$)~arcmin. The primary beam full width at half maximum (FWHM) at 154~MHz is $\approx 27\deg$.

The first year of GLEAM observations used for GLEAM Exgal were conducted between August 2013 and June 2014. The whole sky south of Dec~$+30^\circ$ was surveyed in four week-long campaigns of meridian drift scans \citep{bernardi2013,hurleywalker2014} to obtain overlapping coverage in RA. Observations were carried out at night to avoid contamination from the Sun. The sky was divided into seven Dec strips ($-72\deg$, $-55\deg$, $-40\deg$, $-27\deg$, $-13\deg$, $+2\deg$ and $+18\deg$) and one Dec strip was covered in a given night. 

Each night's observing run was broken into a series of 2-minute scans in five frequency bands of bandwidth 30.72~MHz centred at 87.7, 118.4, 154.2, 185.0 and 215.7~MHz (hereafter 88, 118, 154, 185 and 216~MHz), cycling through the five frequency bands in 10~minutes. Frequencies between 134 and 139~MHz were avoided due to Orbcomm satellite interference. A strong calibrator source was observed in the five frequency bands at the beginning of the observing run. The frequency and time resolution of the correlator output were 40~kHz and 0.5~s, respectively. More details on the observing strategy can be found in \cite{wayth2015}.

In the second year of GLEAM observations conducted between August 2014 and July 2015, twice the amount of observing time was spent surveying the same area of sky at the same frequencies. The observing strategy was adjusted as follows:
\begin{enumerate}
\item[(1)] The observations were divided into eight week-long drift scan campaigns, alternating between an hour angle of +1 and --1. This served to increase the effective $(u,v)$ coverage for each patch of the sky in the final mosaic, and to observe some fields when the brightest, complex radio sources in the sky (the so-called `A-team' sources) were below the horizon rather than in a sidelobe.
\item[(2)] The frequency resolution of the correlator output was set to 10~kHz. The higher frequency resolution was chosen to increase the usefulness of the dataset for spectral line and polarisation science. The time resolution was set to 2~s to retain the overall data rate. A time resolution of 2~s does not lead to significant time-average smearing with the longest baselines in the MWA Phase I array.
\end{enumerate}

\begin{table}
\centering
\caption{GLEAM SGP observing parameters.}
\label{tab:observing_parameters}
\begin{tabular}{@{} c c c c c}
\hline
Date & Year of & RA & Dec & Hour \\
& observation & range (h) & (deg) & angle \\
\hline
2013-08-10 & 1 & $21-3.5$ & $-27$ & 0 \\
2013-08-22 & 1 & $21-3.5$ & $-13$ & 0 \\
2013-08-25 & 1 & $21-3.5$ & $-40$ & 0 \\
2013-11-05 & 1 & $0-5$ & $-13$ & 0 \\
2013-11-06 & 1 & $0-5$ & $-40$ & 0 \\
2013-11-25 & 1 & $0-5$ & $-27$ & 0 \\
2014-06-09 & 1 & $21-22$ & $-27$ & 0 \\
2014-06-10 & 1 & $21-22$ & $-40$ & 0 \\
2014-06-16 & 1 & $21-22$ & $-13$ & 0 \\
2014-08-04 & 2 & $21-1$ & $-27$ & -1 \\
2014-08-05 & 2 & $21-1$ & $-40$ & -1 \\
2014-08-08 & 2 & $21-1$ & $-13$ & -1 \\
2014-09-15 & 2 & $21-4.5$ & $-27$ & +1 \\
2014-09-16 & 2 & $21-4.5$ & $-40$ & +1 \\
2014-09-19 & 2 & $21-4.5$ & $-13$ & +1 \\
2014-10-27 & 2 & $21.5-5$ & $-27$ & -1 \\
2014-10-28 & 2 & $21.5-5$ & $-40$ & -1 \\
2014-10-31 & 2 & $21.5-5$ & $-13$ & -1 \\
2014-12-17 & 2 & $1-5$ & $-40$ & +1 \\
2014-12-20 & 2 & $1-5$ & $-13$ & +1\\
\hline
\end{tabular}
\end{table}
 
\section{Data reduction}\label{Data reduction}

In this paper, we process a subset of the GLEAM year 1 and 2 data at Decs $-40\deg$, $-27\deg$ and $-13\deg$, and in the RA range $21-5~\mathrm{h}$, using a similar procedure to that employed by \cite{hurleywalker2017}. The level of sidelobe contamination from the Galaxy and A-team sources is relatively low in this region of sky, which is centred on the SGP ($00\h 51\m -27\deg 08'$). The SGP also transits through the MWA zenith ($\approx -27\deg$), where the primary beam has the highest sensitivity. Table~\ref{tab:observing_parameters} lists the observations used for GLEAM SGP.

We do not process the data from the lowest frequency band (72--103~MHz) due to calibration errors associated with Fornax A and Pictor A entering the primary beam sidelobes. However, given that GLEAM Exgal is limited by classical confusion in the lowest frequency band, no significant improvement in the sensitivity is expected from combining the GLEAM year 1 and 2 data. The rms noise achieved in the lowest frequency band of GLEAM Exgal is $\approx 40$~mJy/beam while \cite{franzen2019} estimate the classical confusion noise to be 30~mJy/beam. In the final source catalogue, we include flux densities extracted from the GLEAM Exgal mosaics below 100~MHz, as explained in Section~\ref{Source finding and cataloguing}.

The full GLEAM Exgal data reduction procedure is described in detail in \cite{hurleywalker2017}. In this section, we summarise the main calibration and imaging steps used in GLEAM Exgal, describe the changes made to the data reduction process in GLEAM SGP, and compare the sensitivity and dynamic range in the final GLEAM Exgal and SGP mosaics.

\subsection{Summary of original GLEAM Exgal data reduction}\label{Summary of GLEAM Exgal data reduction}

The raw visibility data from each 2-min snapshot observation were pre-processed using \textsc{cotter} \citep{offringa2015}: data affected by radio frequency interference (RFI) were flagged and the data were averaged to a time resolution of 4~s and a frequency resolution of 40~kHz. For each night's observing run, antenna amplitude and phase solutions were derived for a source calibrator observation at the beginning of the observing run and applied to the entire night of drift scan data. The calibration was performed using the Common Astronomy Software Applications (CASA\footnote{http://casa.nrao.edu/}) task \textsc{bandpass}.

The snapshot data were imaged and self-calibrated using the \textsc{wsclean} imager \citep{offringa2014}, which corrects for wide-field \textit{w}-term effects, and the full-Jones \textsc{mitchcal} algorithm developed for MWA calibration \citep{offringa2016b}. At this stage, the 30.72~MHz bandwidth of the data was divided into narrower sub-bands of 7.68~MHz. Images of 7.68~MHz bandwidth at 20 frequencies distributed continuously between 72 and 231~MHz, but avoiding 134--139~MHz, were created.

The Molonglo Reference Catalogue \citep[MRC;][]{large1981,large1991} was used to set a basic flux scale for the snapshot images. This served to remove any residual RA- (time-) dependent flux scale errors from the drift scans. Astrometric calibration was performed in the image plane using \textsc{fits\textunderscore warp} \citep{hurleywalker2018}: the MRC and NRAO VLA Sky Survey \citep[NVSS;][]{condon1998} catalogue were used to correct source position offsets introduced in the snapshot images due to ionospheric distortions.

At each frequency, the snapshot images were corrected for the primary beam using the analytical primary beam model of \cite{sutinjo2015} and mosaicked together. It was necessary to correct for residual Dec-dependent errors in the flux scale due to errors in the adopted primary beam model. This was done by comparing the measured flux densities of sources in the mosaic with their radio spectra as predicted by three catalogues: the VLA Low-Frequency Sky Survey redux \citep[VLSSr;][]{lane2014} at 74~MHz, MRC at 408~MHz and NVSS at 1.4~GHz. A map tracing the variation of the point spread function (PSF) across the mosaic was generated using sources known to be unresolved in higher resolution radio surveys. The PSF map, which describes the apparent blurring of the PSF due to ionospheric smearing, was taken into account when measuring source sizes and integrated flux densities in the mosaics.

The final image products consist of 20 Stokes $I$ 7.68-MHz sub-band mosaics spanning 72--231~MHz as well as a deep wide-band mosaic covering 170--231~MHz, formed by combining the eight highest frequency sub-band mosaics. The wide-band image centred at 200~MHz was used for source detection. Away from the northern and southern edges of the survey at $-72\deg \leq \mathrm{Dec} < 18.5\deg$, the sensitivity is $\approx 10$~mJy/beam and the angular resolution $\approx 2.3$~arcmin.

\subsection{Improvements to data reduction procedure in GLEAM SGP}\label{Changes to data reduction in GLEAM SGP}

\subsubsection{Pre-processing}\label{Pre-processing}

In a single night of GLEAM observing, roughly 10~TB of raw visibility data are generated. Using Dysco compression, \cite{offringa2016a} showed that MWA data with typical time and frequency resolutions used in processing can be compressed by a factor of $\approx 4$ with less than $\approx 1$ per cent added system noise. In order to reduce the volume of visibility data and to speed up processing, after the RFI flagging and averaging steps, the $(u,v)$ data are compressed using Dysco.

\subsubsection{Calibration}\label{Calibration}

We do not use the calibrator source observations to calibrate the snapshot data. Instead, we rely on the GLEAM Exgal catalogue as a sky model to calibrate the data.

For each GLEAM SGP snapshot, the sky model is constructed as follows. We select all sources in the catalogue which lie in the main lobe of the snapshot's primary beam (out to a radius of $\approx 35\deg$ at 118~MHz and $\approx 20\deg$ at 216~MHz). The catalogue contains spectral indices, $\alpha$, and 200~MHz flux densities, $S_{200}$, derived from power-law fits to the sub-band flux densities between 76 and 227~MHz. Where possible, we use $\alpha$ and $S_{200}$ to derive the integrated flux densities, $S_{\nu_{\mathrm{c}}}$, of the sources in the model at the central frequency, $\nu_{\mathrm{c}}$, of the snapshot. There is no measurement of the spectral index for 25 per cent of the sources in the catalogue; the sources with missing spectral indices are mostly the fainter ones for which there was insufficient signal-to-noise to make a reliable measurement. If $\alpha$ and $S_{200}$ are not available in the catalogue, we derive $S_{\nu_{\mathrm{c}}}$ by taking the sub-band flux density closest in frequency to $\nu_{\mathrm{c}}$ and extrapolating it to $\nu_{\mathrm{c}}$ with $\alpha = -0.8$, the typical spectral index of sources seen in GLEAM Exgal \citep{hurleywalker2017}.

The morphology of the sources in GLEAM Exgal is best characterised using the wide-band mosaic covering 170--231~MHz as it has the best sensitivity and resolution. The GLEAM Exgal catalogue provides the major axis, $a_{\mathrm{wide}}$, minor axis, $b_{\mathrm{wide}}$, and position angle, $\theta_{\mathrm{wide}}$, of each source, derived from Gaussian fitting in the wide-band mosaic. The ratio of the integrated to peak flux density, $R$, in the wide-band mosaic can be used to distinguish between point-like and extended sources.

We use our estimate of $S_{\nu_{\mathrm{c}}}$, and the source position and $\alpha$ from the GLEAM Exgal catalogue, to characterise the sources in the model. If $\alpha$ is not available in the catalogue, we set the spectral index to --0.8. Sources in the sky model with $R < 1.2$ are modelled as point sources. The remaining sources are modelled as Gaussian components. To characterise the Gaussian components, we use, in addition, $a_{\mathrm{wide}}$, $b_{\mathrm{wide}}$ and $\theta_{\mathrm{wide}}$. Finally, an apparent flux density cut of 100~mJy is applied to limit the processing time needed to generate the model visibilities, which is proportional to the number of components in the sky model. The final number of sources present in the sky model varies between $\approx 10,000$ at the lowest frequency (118~MHz) to $\approx 1,500$ at the highest frequency (216~MHz).

Calibration is performed using \textsc{mitchcal}, scaling each component in the sky model by a model of the primary beam by \cite{sokolowski2017}. This primary beam model is more accurate than that of \cite{sutinjo2015} as every single dipole in the MWA tile is simulated separately, taking into account all mutual coupling, ground screen and soil effects. As in GLEAM Exgal, baselines shorter than 60~m are not used to perform the calibration due to contamination from expected large-scale Galactic emission.

\cite{hurleywalker2017} removed areas within 10~arcmin of the following A-team sources from the GLEAM Exgal catalogue due to the difficulty in calibrating and imaging them at low frequencies: Centaurus A, Taurus A, Hercules A, Hydra A, Pictor A, Virgo A, Orion A, Fornax A and Tau A. The calibration solutions may therefore be significantly affected by the absence of these sources in the sky model. Under good ionospheric conditions, the amplitude and phase solutions are expected to vary smoothly as a function of frequency and to remain stable during the night's observing run. From visual inspection of the calibration solutions, we find that $\approx 25$ per cent of the snapshots are significantly affected. Most of these snapshots are located within $\approx 20$~deg of Fornax A and/or Pictor A, while a smaller number of them are located in a sidelobe of Cygnus A.

A different procedure is used to calibrate the data from these affected snapshots: we apply good calibration solutions derived from the GLEAM Exgal sky model for other snapshots at the same frequency and separated by no more than 2~hours in time. We then perform self-calibration following the same procedure as adopted in GLEAM Exgal: a sky model is constructed from the observation itself and the model is used to further improve the calibration solutions (see \citealp{hurleywalker2017} for details). For the remaining snapshots, this additional self-calibration loop is not performed as it is not found to significantly improve the calibration solutions.

\subsubsection{Snapshot imaging}\label{Snapshot imaging}

Most of the processing time for GLEAM Exgal was spent on imaging the large number of snapshots within the survey. The GLEAM Exgal snapshots were imaged using \textsc{wsclean} v1.10. The image size was set to $4000 \times 4000$ pixels to cover the primary beam down to the $\approx 10$ per cent level and a robust weighting of $-1.0$ \citep[close to uniform][]{briggs1995} was used. An initial estimate of the rms noise, $\sigma$, was made by imaging the snapshot down to the first negative CLEAN component. The snapshot was then re-imaged down to a CLEAN threshold of 3$\sigma$.

We image the $\approx 2400$ 30.72~MHz band snapshots within GLEAM SGP using \textsc{wsclean} v2.5, which is more efficient for large images thanks to the implementation of the Clark CLEAN algorithm \citep{clark1980}. In minor CLEAN cycles, only the central portion of the synthesised beam is used to subtract the CLEAN components from the image and only the largest CLEAN components are searched for. This is sufficient to find the CLEAN components providing that the synthesised beam is well behaved. We use the same robust weighting of --1.0 but set the minimum $(u,v)$ distance to $30\lambda$ and taper the weights with a Tukey transition of size $15\lambda$. This removes structure larger than $\approx 1.5$~deg prior to imaging, serving to reduce the sidelobe contamination from the Galaxy. We initially CLEAN the entire image down to a threshold of 3$\sigma$ and simultaneously create a mask containing all found components. We then continue CLEANing with the constructed mask down to a deeper threshold of 1$\sigma$. This CLEANing procedure is fully automated and allows stuctures to be CLEANed down to the noise level around detected sources \citep{offringa2017}.

The flux scale of the snapshot images is expected to be too low as a result of flux uncaptured in the GLEAM Exgal sky model (as mentioned above, all sources with $S_{\nu_{\mathrm{c}},\mathrm{app}} < 100$~mJy were removed from the model). In GLEAM Exgal, the MRC was used to set a basic flux scale for the snapshot images; a selected sample of sources in the snapshot image were cross-matched with MRC and the measured flux densities were compared with those predicted from MRC. We adopt the same procedure to correct the flux scale of the snapshot images in GLEAM SGP except that we use the GLEAM Exgal catalogue to set the flux scale.

\subsubsection{Mosaicking}\label{Mosaicking}

We follow the exact same procedure as in GLEAM Exgal to combine the GLEAM year 1 and 2 snapshots at each frequency into mosaics and apply the Dec-dependent flux scale correction, with the following exceptions:

\begin{enumerate}
\item[(1)] The synthesised beam size of the snapshots at each frequency is found to vary by up to $\approx 10$ per cent due to slight changes in the $(u,v)$ coverage of the observations. Before mosaicking, each snapshot is convolved with a Gaussian to obtain an identical synthesised beam at each frequency.
\item[(2)] In order to improve the accuracy of the flux scale, we use the primary beam model by \cite{sokolowski2017} in the mosaicking step.
\item[(3)] The mosaicking process results in a set of 16 images between 107 and 227~MHz, each with a bandwidth of 7.68~MHz. In GLEAM Exgal, the most sensitive combined image was obtained by combining the eight highest frequency sub-band mosaics at 170--231~MHz. In GLEAM SGP, we find that combining the four highest frequency sub-band mosaics at 200--231~MHz results in a better compromise between sensitivity and resolution.
\end{enumerate}

As mentioned in Section~\ref{Summary of GLEAM Exgal data reduction}, ionospheric perturbations can cause sources to be slightly smeared out in the mosaicked images. We therefore generate maps of the spatial variation of the PSF across each of the mosaics using the method adopted in GLEAM Exgal. The mean $\pm$ standard deviation of the major and minor axes of the PSF in the wide-band mosaic are ($2.45 \pm 0.05$)~arcmin and ($1.98 \pm 0.05$)~arcmin, respectively.

The wide-band mosaic centred at 216~MHz is shown in Fig.~\ref{fig:mosaic}. We use this image for source detection, as described in Section~\ref{Source finding and cataloguing}. 

\begin{landscape}
\centering
\begin{figure}
\includegraphics[scale=0.85, angle=270, trim=3.5cm 0cm 3cm 0cm]{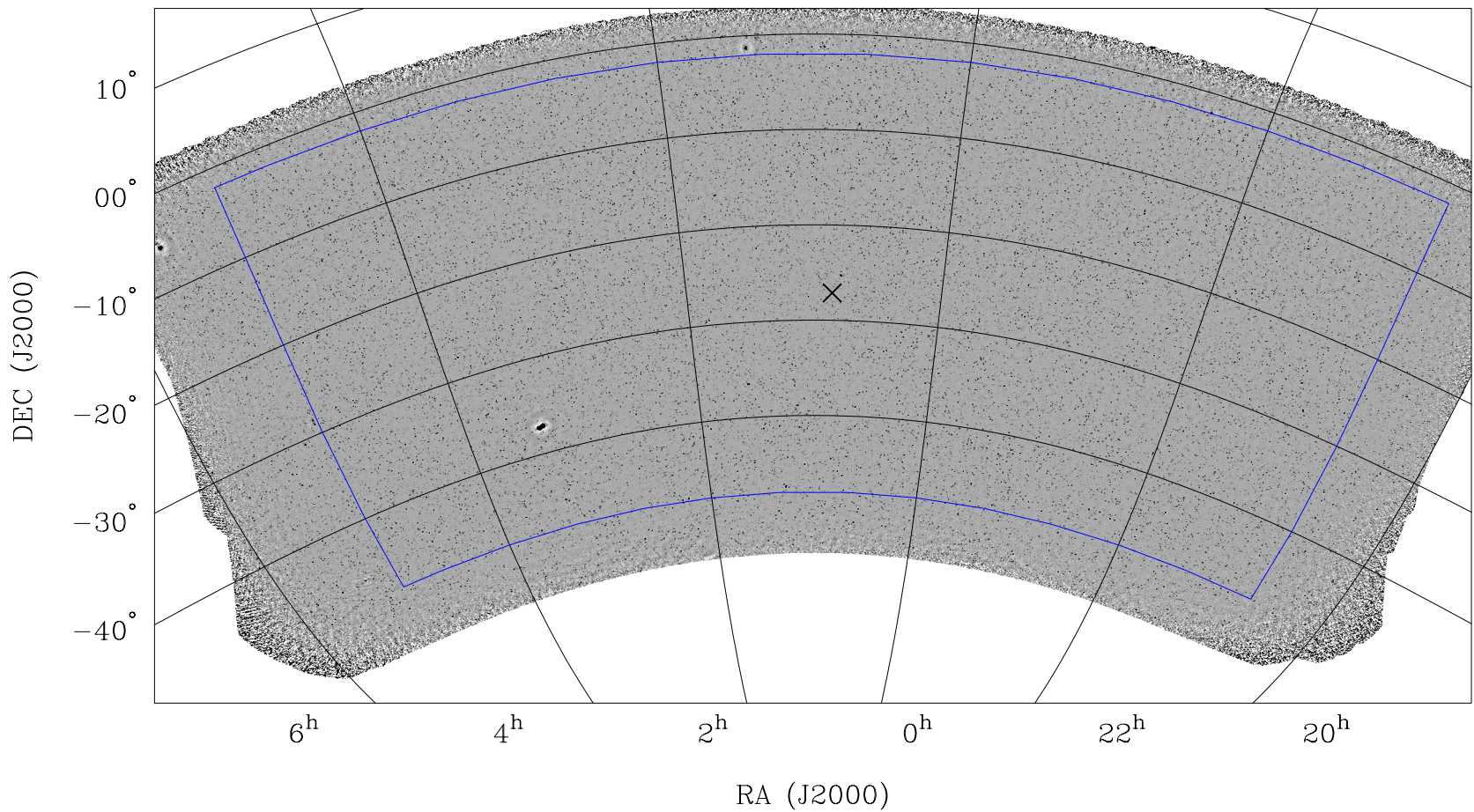}
\caption{The GLEAM SGP wide-band mosaic centred at 216~MHz. The grey-scale is linear and runs from --50 to 100~mJy/beam. The blue line indicates the catalogue boundary of the mosaic, chosen as described in Section~\ref{Improvement in sensitivity}. The black cross marks the SGP.}
\label{fig:mosaic}
\end{figure}
\end{landscape}

\subsection{Improvement in sensitivity}\label{Improvement in sensitivity}

We compare the rms noise in the GLEAM Exgal and SGP wide-band mosaics. We create rms noise maps of the two mosaics using \textsc{BANE} \citep{hancock2018}. Using these noise maps, we define an area of sky where the rms noise is lower in GLEAM SGP: $20\h 40\m < \mathrm{RA} < 05\h 04\m$ and $-48\deg < \mathrm{Dec} < -2\deg$. In this 5,113$~\mathrm{deg}^2$ area of sky, hereafter referred to as the GLEAM SGP region, the mean rms noise in GLEAM SGP (4.7~mJy/beam) is $\approx 40$ per cent lower than that in GLEAM Exgal (7.6~mJy/beam). The best improvement in the rms in GLEAM SGP is by a factor of $\approx 3$ and is seen in areas surrounding bright sources due to the reduction in image artefacts.

\begin{table*}
\centering
\caption{Comparison of the GLEAM Exgal and SGP survey properties in the GLEAM SGP region ($20\h 40\m < \mathrm{RA} < 05\h 04\m$ and $-48\deg < \mathrm{Dec} < -2\deg$). Values are given as the mean $\pm$ the standard deviation. The statistics shown are derived from the GLEAM Exgal wide-band mosaic at 170--231~MHz and the GLEAM SGP wide-band mosaic at 200--231~MHz. The RA and Dec astrometric offsets show the degree to which the source positions agree with NVSS and SUMSS; the RA offset is given by $\mathrm{RA}_{\mathrm{NVSS/SUMSS}} - \mathrm{RA}_{\mathrm{GLEAM}}$ and the Dec offset by $\mathrm{Dec}_{\mathrm{NVSS/SUMSS}} - \mathrm{Dec}_{\mathrm{GLEAM}}$. The external flux scale error applies to all frequencies and shows the degree to which the source flux densities agree with other published surveys. The internal flux scale error also applies to all frequencies and shows the internal consistency of the flux scale.}
\label{tab:survey_properties}
\begin{tabular}{@{} c c c}
\hline
Property & GLEAM Exgal & GLEAM SGP \\
\hline
Number of sources & 85,981 & 108,851 \\
RA astrometric offset (arcsec) & $-0.2 \pm 3.3$ & $-0.5 \pm 3.3$ \\
Dec astrometric offset (arcsec) & $-1.6 \pm 3.3$ & $-1.0 \pm 2.9$ \\
External flux scale error (\%) & 8 & 8 \\
Internal flux scale error (\%) & 2 & 2 \\
RMS (mJy/beam) & $7.6 \pm 1.8$ & $4.7 \pm 1.6$ \\
PSF major axis (arcsec) & $135 \pm 3$ & $147 \pm 3$ \\
PSF minor axis (arcsec) & $130 \pm 2$ & $119 \pm 3$ \\
\hline
\end{tabular}
\end{table*}

The number of constituent snapshot images covering any patch of sky in the GLEAM SGP mosaic is about three times higher than that in the GLEAM Exgal mosaic. The thermal noise is therefore expected to be $\approx \sqrt{3}$ times lower in GLEAM SGP. No change in the classical confusion noise is expected since the two survey releases have very similar PSF sizes (see Table~\ref{tab:survey_properties}). A reduction in the overall sidelobe levels of un-deconvolved sources is expected in GLEAM SGP due to the better $(u,v)$ coverage and improved deconvolution of the snapshot images. Since sidelobe confusion is the dominant noise contribution in GLEAM Exgal above $\approx 100$~MHz, most of the reduction in the rms noise in GLEAM SGP results from the lower sidelobe confusion noise.

\begin{figure}
\begin{center}
\includegraphics[scale=0.6, trim=0cm 4cm 7.5cm 0cm, angle=270]{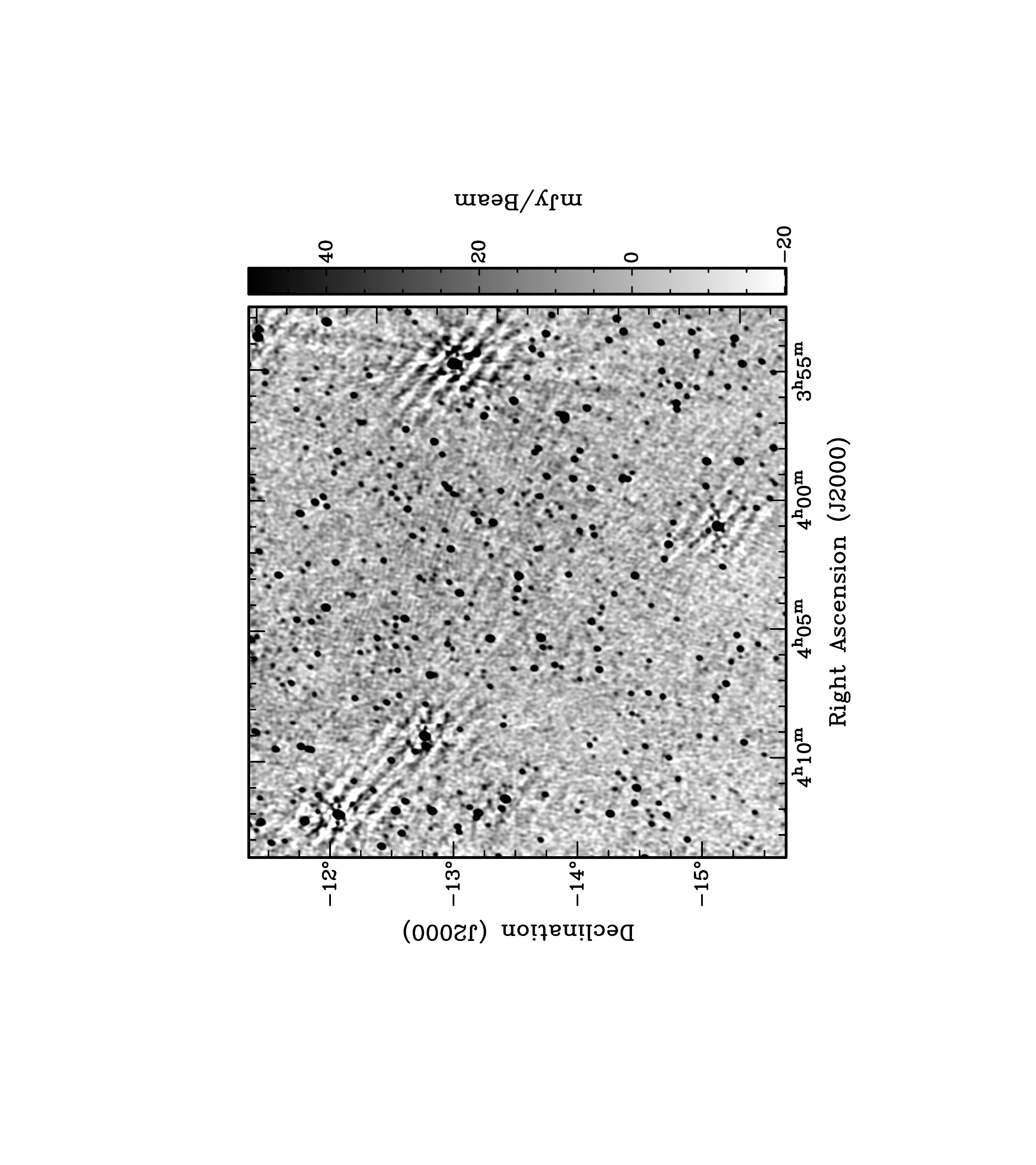}
\includegraphics[scale=0.6, trim=0cm 4cm 3.5cm 0cm, angle=270]{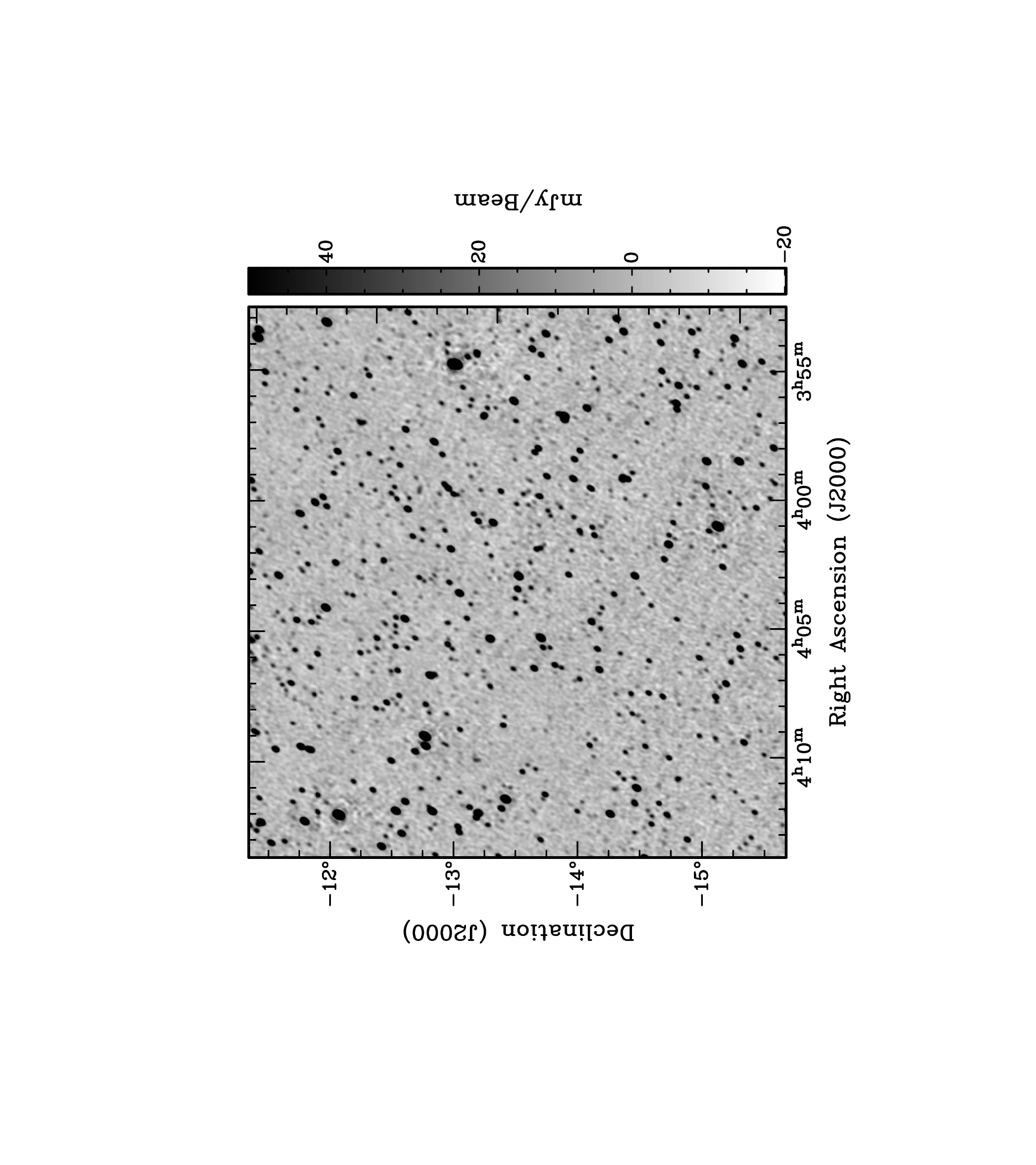}
\caption{Example 25\sqdeg~of sky containing several bright sources from the GLEAM Exgal wide-band mosaic at 170--231~MHz (top) and the GLEAM SGP wide-band mosaic at 200--231~MHz (bottom), highlighting the improvement in the rms in GLEAM SGP. The grey scale is linear and runs from --20 to 50~mJy/beam in both panels.}
\label{fig:Exgal_SGP_cutout_comparison}
\end{center}
\end{figure}

Fig.~\ref{fig:Exgal_SGP_cutout_comparison} compares an example 25\sqdeg~of sky in the GLEAM Exgal and SGP wide-band mosaics. The reduction in image artefacts around bright sources in the GLEAM SGP mosaic likely results from the improved calibration and deconvolution of the snapshots, as well as the larger number of snapshots contributing to any patch of sky. The diffuse Galactic synchrotron emission visible in the GLEAM Exgal mosaic is also largely removed in the GLEAM SGP mosaic as a result of the $(u,v)$ taper applied to the GLEAM SGP data.

\subsection{GLEAM year 1 and 2 mosaics}\label{GLEAM year 1 and 2 mosaics}

We create additional mosaics for the GLEAM year 1 and 2 data by combining the GLEAM year 1 and 2 snapshots at each frequency into mosaics separately.  The GLEAM year 1 data were taken almost entirely over the period August--November 2013 and the GLEAM year 2 data were taken over the period August--December 2014, as shown in Table~\ref{tab:observing_parameters}. These two epochs of data therefore provide a unique opportunity to search for low-frequency variability in the flux density over a large fractional bandwidth on a timescale of approximately one year. The mean rms noise in the GLEAM year 1 and 2 wide-band (200--231~MHz) mosaics within the GLEAM SGP region is 6.5 and 5.5~mJy/beam, respectively.

\section{Source finding and cataloguing}\label{Source finding and cataloguing}

Following \cite{hurleywalker2017}, we perform blind source finding on the wide-band mosaic covering 200--231~MHz to obtain a reference catalogue. We then extract the flux densities of each source within the reference catalogue in the sub-band images.

The source finding on the wide-band mosaic is performed as follows. We first use \textsc{bane} to estimate the background emission and rms noise across the mosaic. Next, we run \textsc{aegean} \citep{hancock2012,hancock2018} on the mosaic using a detection threshold of 5 $\times$ the local rms. The spatial variation of the PSF is taken into account using the PSF map. Each detected source is characterised by \textsc{aegean} as an elliptical Gaussian component. Six parameters are fitted for each component: the peak RA and Dec, peak flux density, major and minor axes, and position angle.

In order to remove potential spurious detections, all sources within 0.5~deg of Fornax A are discarded. After removing these sources, the total number of sources detected in the GLEAM SGP region is 108,851. In comparison, the GLEAM Exgal catalogue contains 85,981 sources in the GLEAM SGP region.

For each source in the reference catalogue, we extract the flux density in each of the 16 sub-band images between 107 and 227~MHz using the `priorised fitting' mode of \textsc{aegean}. The expected shape of the source in the sub-band image is derived by \textsc{aegean} given its shape in the wide-band image, and the local PSFs from the wide-band and sub-band images. A fit is performed for the peak flux density of each source; the position and newly-determined shape of the source are not allowed to vary.

We use the four lowest frequency sub-band mosaics from GLEAM Exgal to extract additional flux density measurements between 76 and 99~MHz for all sources within the reference catalogue, via priorised fitting.

The advantage of this priorised fitting approach is that it provides measurements for all sources in the reference catalogue across the full frequency range without having to rely on position-based cross-matching of catalogues. The positions and morphologies of the sources are most precisely determined in the wide-band image which has the best resolution and sensitivity, and this information is used to constrain the source flux densities in the sub-band images. Since the flux densities in the sub-band images are not extracted via blind source finding, no signal-to-noise ratio (SNR) threshold is applied to the flux densities from the sub-band images. Sources may therefore be detected well below 5$\sigma$ in the sub-band images, or even have negative flux densities.

\subsection{Spectral indices}\label{Spectral indices}

We calculate the spectral indices between 76 and 227~MHz of the GLEAM SGP sources from the 20 sub-band flux densities. For the spectral index of a source to be calculated, it must have a positive flux density measurement in each of the 20 sub-bands; this is the case for 77 per cent of the sources in the catalogue. For these 83,328 sources, we calculate $\alpha$ using a weighted least-squares approach. The flux density error in each sub-band is taken as the sum in quadrature of the Gaussian fitting error (as calculated by \textsc{aegean}) and a calibration error of 2 per cent. We estimate the internal flux calibration error to be 2 per cent from the reduced $\chi^2$ statistic for bright sources: at high flux densities ($S_{216~\mathrm{MHz}} \gtrsim 1~\mathrm{Jy}$), the median reduced $\chi^2$ value should be close to 1.0 if the internal flux calibration error has been well estimated. We find this to be the case when using an internal flux calibration error of 2 per cent.

In addition to the spectral index, the fitted 200-MHz flux density and reduced $\chi^2$ value from the least-squares fitting are provided in the catalogue. The reduced $\chi^2$ value can be used to assess the quality of the fitted spectral index and 200-MHz flux density: for 18 degrees of freedom, $P$(reduced $\chi^2 > 1.93$)~$<1$\% and $P$(reduced $\chi^2 > 2.35$)~$<0.1$\%. 

\begin{figure}
\begin{center}
\includegraphics[scale=0.5, trim=0cm 0cm 0cm 0cm]{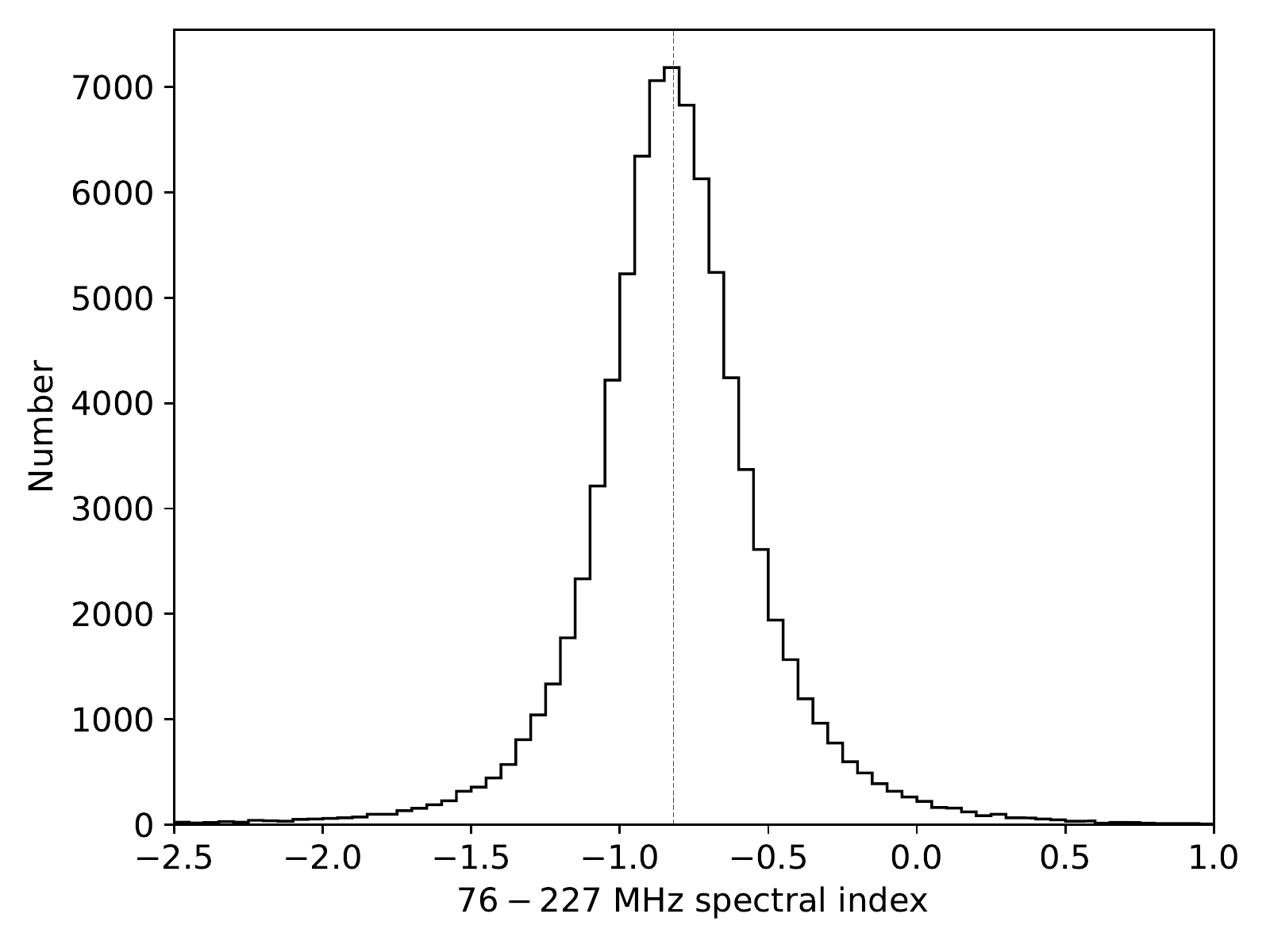}
\caption{The spectral index distribution of the GLEAM SGP sources measured using the 20 sub-band flux densities between 76 and 227~MHz. Only sources with reduced $\chi^2 < 1.93$ and $\delta \alpha < 0.5$ are included. The dashed vertical line shows the median spectral index (--0.82).}
\label{fig:plot_alpha_dist}
\end{center}
\end{figure}

Of the 83,328 sources with measured spectral indices, 81,892 (98.3 per cent) have reduced $\chi^2 < 1.93$ and spectral index error, $\delta \alpha < 0.5$. Fig.~\ref{fig:plot_alpha_dist} shows the spectral index distribution of these sources. The mean and median spectral indices are --0.81 and --0.82, respectively.

A tiny fraction of the sources with reduced $\chi^2 < 1.93$ and $\delta \alpha < 0.5$ have spectral indices which are implausibly steep; 25 sources have $\alpha < -3$ and the minimum spectral index is --4.16. In the priorised fitting, the lowest sub-band flux densities of these sources are significantly overestimated due to severe source confusion, causing their spectral indices to be too steep. We caution that if a source lies within $\approx 5$~arcmin (the beam size in the lowest frequency sub-band) from another source whose flux density is at least $\approx 10$ times higher in the wide-band image, its measured spectral index may be significantly affected by confusion.

\subsection{Extended sources}\label{Extended sources}

\begin{figure}
\begin{center}
\includegraphics[scale=0.5, trim=1cm 0cm 0cm 0cm]{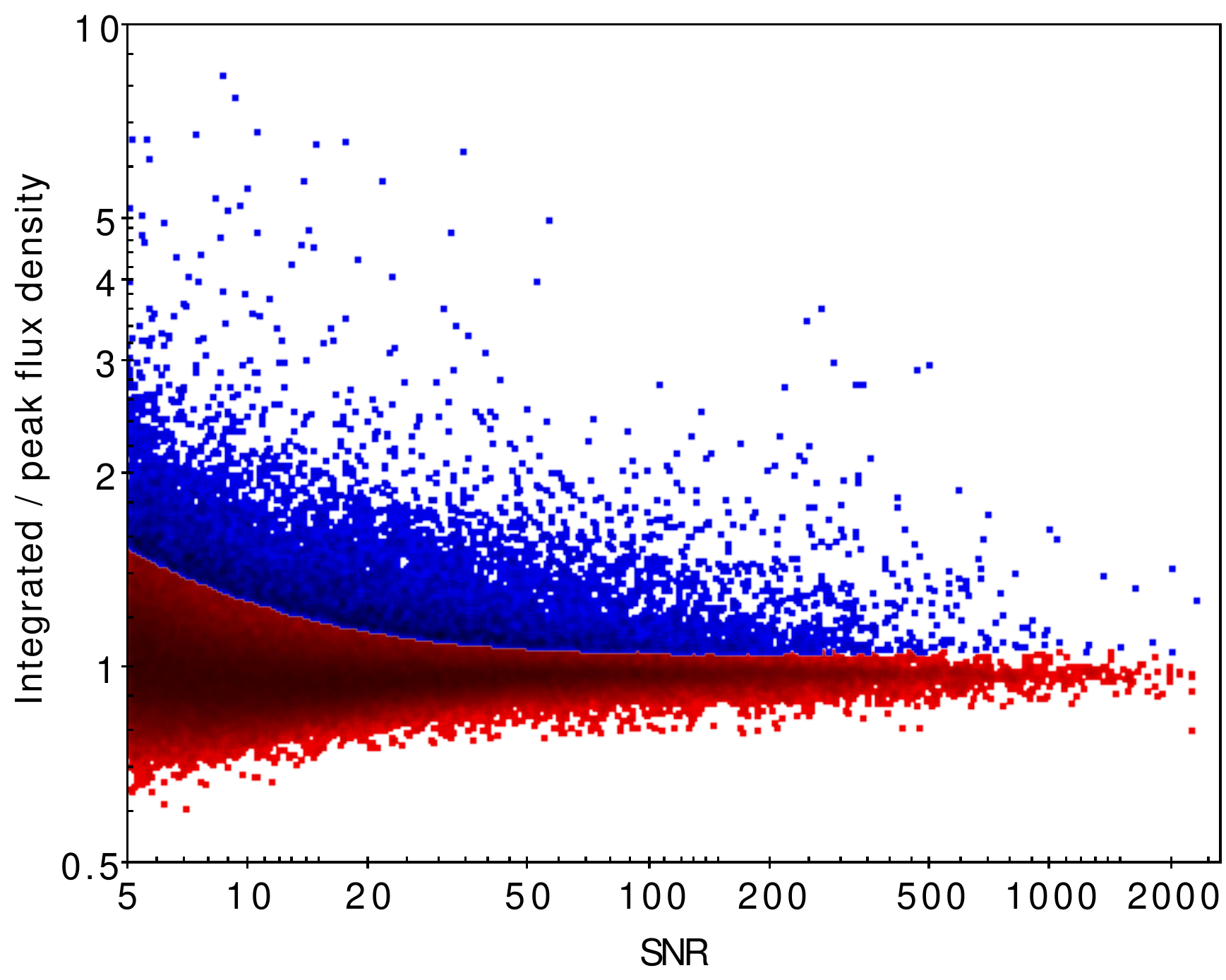}
\caption{The ratio of the integrated to peak flux density as a function of the SNR for all GLEAM SGP sources detected in the wide-band image. Sources classified as point-like are shown in red and as extended in blue.}
\label{fig:plot_R_versus_SNR}
\end{center}
\end{figure}

A common method for classifying sources as point-like or extended is through the ratio of their integrated to peak flux densities. Fig.~\ref{fig:plot_R_versus_SNR} shows the ratio of the integrated to peak flux density in the wide-band image, $\frac{S}{S_{\mathrm{p}}}$, as a function of the SNR, for all GLEAM SGP sources. Instances where $\frac{S}{S_{\mathrm{p}}} < 1.0$ are due to uncertainties in the source size introduced by the image noise and calibration errors.

We use the following relation in the wide-band image to separate point-like from extended sources taking these uncertainties into account: 
\begin{equation}\label{eqn:extended_relation}
\frac{S}{S_{\mathrm{p}}} > 1.0 + a \sqrt{ \left(\frac {\sigma_{\mathrm{local}}} {S_{\mathrm{p}}} \right)^{2} + \epsilon^{2}}\ ,
\end{equation}
where $\sigma_{\mathrm{local}}$ is the local rms noise and $\epsilon$ the internal flux scale error. We set $\epsilon = 0.02$ based on the analysis carried out in Section~\ref{Spectral indices}. Following a similar approach to that of \cite{butler2018}, we set the value of $a$ such that 95 per cent of the sources with $\frac{S}{S_{\mathrm{p}}} < 1.0$ lie above the curve defined by
\begin{equation}\label{eqn:extended_relation_mirror}
\frac{S}{S_{\mathrm{p}}} = 1.0 - a \sqrt{ \left(\frac {\sigma_{\mathrm{local}}} {S_{\mathrm{p}}} \right)^{2} + \epsilon^{2}}\ .
\end{equation}

The resulting value of $a$ is 2.77. At high SNR where calibration errors dominate, $\frac{S}{S_{\mathrm{p}}} > 1.06$ for a source to be classified as extended; at an SNR of 5, $\frac{S}{S_{\mathrm{p}}} > 1.56$ for a source to be classified as extended.

Using Equation~\ref{eqn:extended_relation}, 8.4 per cent of the GLEAM SGP sources are classified as extended, where the beam size is $\approx 2$~arcmin. These sources are shown in blue in Fig.~\ref{fig:plot_R_versus_SNR} and are flagged in the catalogue. Given the large beam size, a large fraction of these sources are not expected to be genuinely extended, but rather the result of source confusion.

\subsection{Completeness and reliability}\label{Completeness and reliability}

Simulations in the image plane are used to quantify the completeness of the source catalogue. Following the same method as in GLEAM Exgal, we inject artificial point sources with flux densities ranging between 15 and 300~mJy into the wide-band mosaic used for source detection. We then create maps tracing the variation of the completeness across the sky at the various flux density levels. The completeness at any pixel position is given by the fraction of simulated sources that are detected above $5\sigma$ in a circle of radius $6\deg$ centred on the pixel. Full details of the procedure are explained in \cite{hurleywalker2017}. The completeness maps in \textsc{fits} format can be obtained from the supplementary material.

\begin{figure}
\begin{center}
\includegraphics[scale=0.5, trim=0cm 0cm 0cm 0cm]{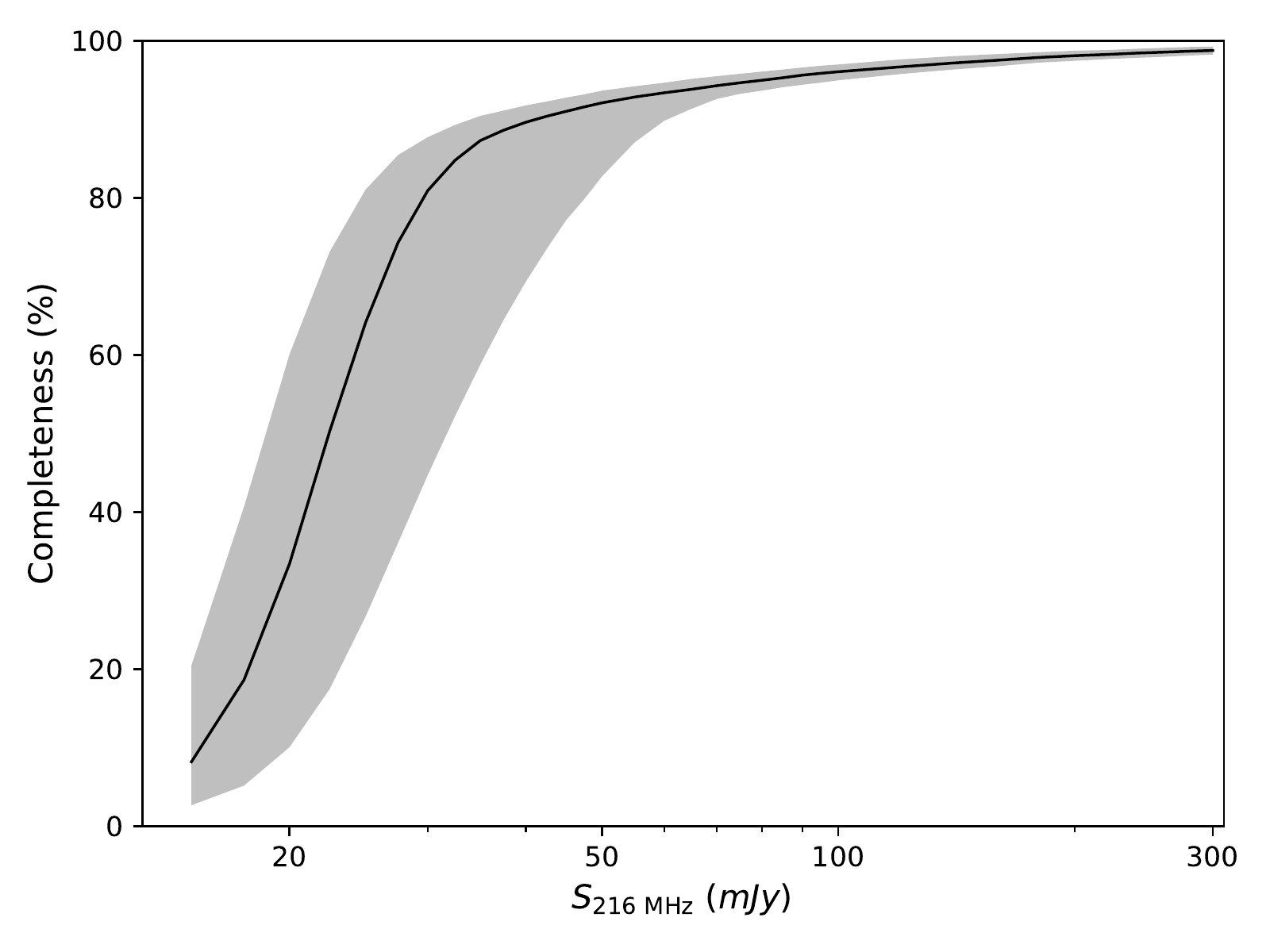}
\caption{Estimated completeness across the GLEAM SGP region as a function of $S_{216~\mathrm{MHz}}$. The black curve shows the median completeness and the shaded area the 10--90 percentile range.}
\label{fig:completeness}
\end{center}
\end{figure}

The black curve in Fig.~\ref{fig:completeness} shows the median completeness across the GLEAM SGP region as a function of $S_{216~\mathrm{MHz}}$; the shaded area indicates the 10--90 percentile range. The completeness of the source catalogue is estimated to be 50\% at $\approx 25$~mJy, rising to 90\% at $\approx 50$~mJy.

In order to assess the reliability of the source catalogue, we use \textsc{Aegean} to only search for sources with flux densities below $-5\sigma$ in the wide-band image. In total, 7 sources with negative peaks below $-5\sigma$ are detected in the GLEAM SGP region. Assuming the noise distribution is close to symmetric about zero, we can expect to find an approximately equal number of false positive sources in the same area. The total number of sources detected above $5\sigma$ is 108,851. We therefore estimate the catalogue reliability to be:
\begin{equation}
1.0 - \frac{7}{108,851} = 99.994\% \, .
\end{equation}

\subsection{Catalogue validation}\label{Catalogue validation}

\subsubsection{Positional accuracy}\label{Positional accuracy}

The positions of the GLEAM SGP sources are extracted from the wide-band mosaic at 200--231~MHz, used for source detection. In order to verify the astrometry, we cross-match our catalogue with the higher-resolution NVSS catalogue at $\delta \geq -39.5\deg$ and the higher-resolution Sydney University Molonglo Sky Survey \citep[SUMSS;][]{mauch2003} catalogue at $\delta < -39.5\deg$; NVSS has an angular resolution of 45~arcsec and SUMSS has an angular resolution of $45''\times45'' \operatorname{cosec} |\delta|$. We only include unresolved ($S_{\mathrm{int}} / S_{\mathrm{pk}} < 1.1$ in the wide-band mosaic), isolated (no internal match within 10~arcmin) GLEAM SGP sources and isolated (no internal match within 3~arcmin) NVSS and SUMSS sources. RA and Dec offsets are measured with respect to the NVSS and SUMSS positions, which are assumed to be correct. 

\begin{figure}
\begin{center}
\includegraphics[scale=0.5, trim=0cm 0cm 0cm 0cm]{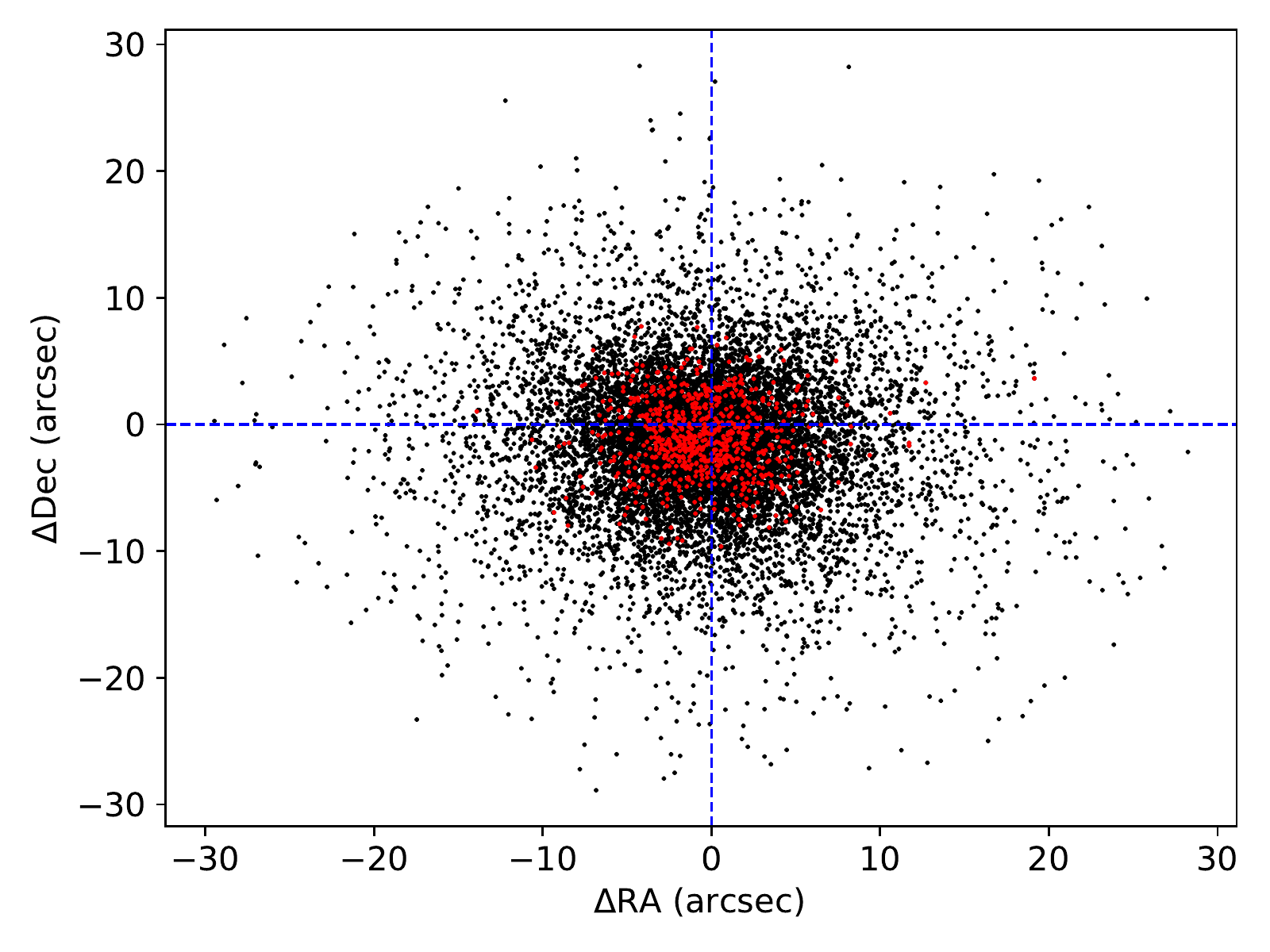}
\caption{RA and Dec offsets for GLEAM SGP sources cross-matched with NVSS or SUMSS, as described in the text. Sources with GLEAM SGP SNRs > 100 are shown in red and the rest of the sources in black.}
\label{fig:astrometry}
\end{center}
\end{figure}

Fig.~\ref{fig:astrometry} shows the RA offset, $\Delta \mathrm{RA} = \mathrm{RA}_{\mathrm{NVSS/SUMSS}} - \mathrm{RA}_{\mathrm{GLEAM}}$, and Dec offset, $\Delta \mathrm{Dec} = \mathrm{Dec}_{\mathrm{NVSS/SUMSS}} - \mathrm{Dec}_{\mathrm{GLEAM}}$, for the 10,741 unresolved, isolated sources in common between GLEAM SGP and NVSS/SUMSS. Sources with GLEAM SGP signal-to-noise ratios (SNRs) $\geq 100$, for which calibration errors dominate the position uncertainties, are shown in red. We use the position offsets for these 783 high-SNR sources to estimate the calibration errors $\sigma_{\mathrm{RA, cal}}$ and $\sigma_{\mathrm{Dec, cal}}$ in RA and Dec. The rms deviation of $\Delta$RA is 3.3~arcsec and the rms deviation of $\Delta$Dec is 3.1~arcsec.\footnote{The results obtained when cross-matching GLEAM SGP sources with NVSS and SUMSS separately are similar: when cross-matching with NVSS, the rms deviation of $\Delta$RA is 3.2~arcsec and the rms deviation of $\Delta$Dec is 3.2~arcsec. When cross-matching with SUMSS, the rms deviation of $\Delta$RA is 3.6~arcsec and the rms deviation of $\Delta$Dec is 2.7~arcsec.} We therefore set $\sigma_{\mathrm{RA, cal}} = 3.3~\mathrm{arcsec}$ and $\sigma_{\mathrm{Dec, cal}} = 3.1~\mathrm{arcsec}$.

In the GLEAM SGP catalogue, we set the total position errors $\sigma_{\mathrm{RA}}$ and $\sigma_{\mathrm{Dec}}$ in RA and Dec to
\begin{equation}
\sigma_{\mathrm{RA}} = \sqrt{\sigma_{\mathrm{RA,cal}}^2 + \sigma_{\mathrm{RA,fit}}^2}
\end{equation}
\begin{equation}
\sigma_{\mathrm{Dec}} = \sqrt{\sigma_{\mathrm{Dec,cal}}^2 + \sigma_{\mathrm{Dec,fit}}^2} \, ,
\end{equation}
where $\sigma_{\mathrm{RA, fit}}$ and $\sigma_{\mathrm{Dec, fit}}$ are the Gaussian fitting errors calculated by \textsc{aegean}, which account for the image noise.

\subsubsection{Flux scale accuracy}\label{Flux scale accuracy}

As part of the mosaicking procedure, Dec-dependent flux scale corrections were applied by comparing the flux densities of sources in the mosaics with the flux densities predicted from the SEDs formed from VLSSr, MRC and NVSS flux density measurements. We calculate the external flux scale error at each frequency as the standard deviation of a Gaussian fit to the remaining variation in the ratio of the predicted to measured source flux densities. The percentage uncertainties lie between 6.6 and 7.9 per cent. For simplicity, we set the external flux scale error to 8 per cent at all frequencies.

\cite{hale2019} produced an image of the XMM Large-Scale Structure (XMM-LSS) field with LOFAR at 144~MHz. Their image is centred at RA = $02\h 20\m$, Dec = $-04\deg30'$ and covers an area of $\approx 27$\sqdeg. The central rms noise is 280\ujybm and the angular resolution 7.5 by 8.5~arcsec. Most of the area covered by this image lies within the GLEAM SGP region (the northern extremity of the image at $\mathrm{Dec} > -2\deg$ lies outside the GLEAM SGP region). 

Only relatively bright ($S_{216~\mathrm{MHz}} \gtrsim 1~\mathrm{Jy}$) sources were used in the above assessment of the external flux scale error. As a further check of the GLEAM SGP flux scale at lower flux densities, we cross-match our catalogue with the more sensitive and higher resolution catalogue by \citeauthor{hale2019}.

The much smaller synthesised beam size of LOFAR means that any large-scale, diffuse emission is more likely to be resolved out in the LOFAR image. In order to minimise discrepancies between the GLEAM SGP and LOFAR flux densities arising from extended emission resolved out with LOFAR but not with the MWA, we only consider single-component sources with integrated-to-peak flux densities less than 2 in the LOFAR catalogue. We discard sources that lie within 2~arcmin (the approximate synthesised beam size of the GLEAM SGP wide-band image) of another source in the LOFAR image and may therefore be confused in the GLEAM SGP image. We use the GLEAM SGP flux densities from the wide-band image at 216~MHz. We extrapolate the LOFAR flux densities to 216~MHz using the intra-band spectral indices in the GLEAM SGP catalogue; sources which do not have quoted spectral indices in the catalogue are discarded, as well as sources with poorly-fit spectral indices (reduced $\chi^2 < 1.93$) and spectral index errors greater than 0.5. This leaves a total of 82 sources to use for the flux density comparison.

\begin{figure}
\begin{center}
\includegraphics[scale=0.53]{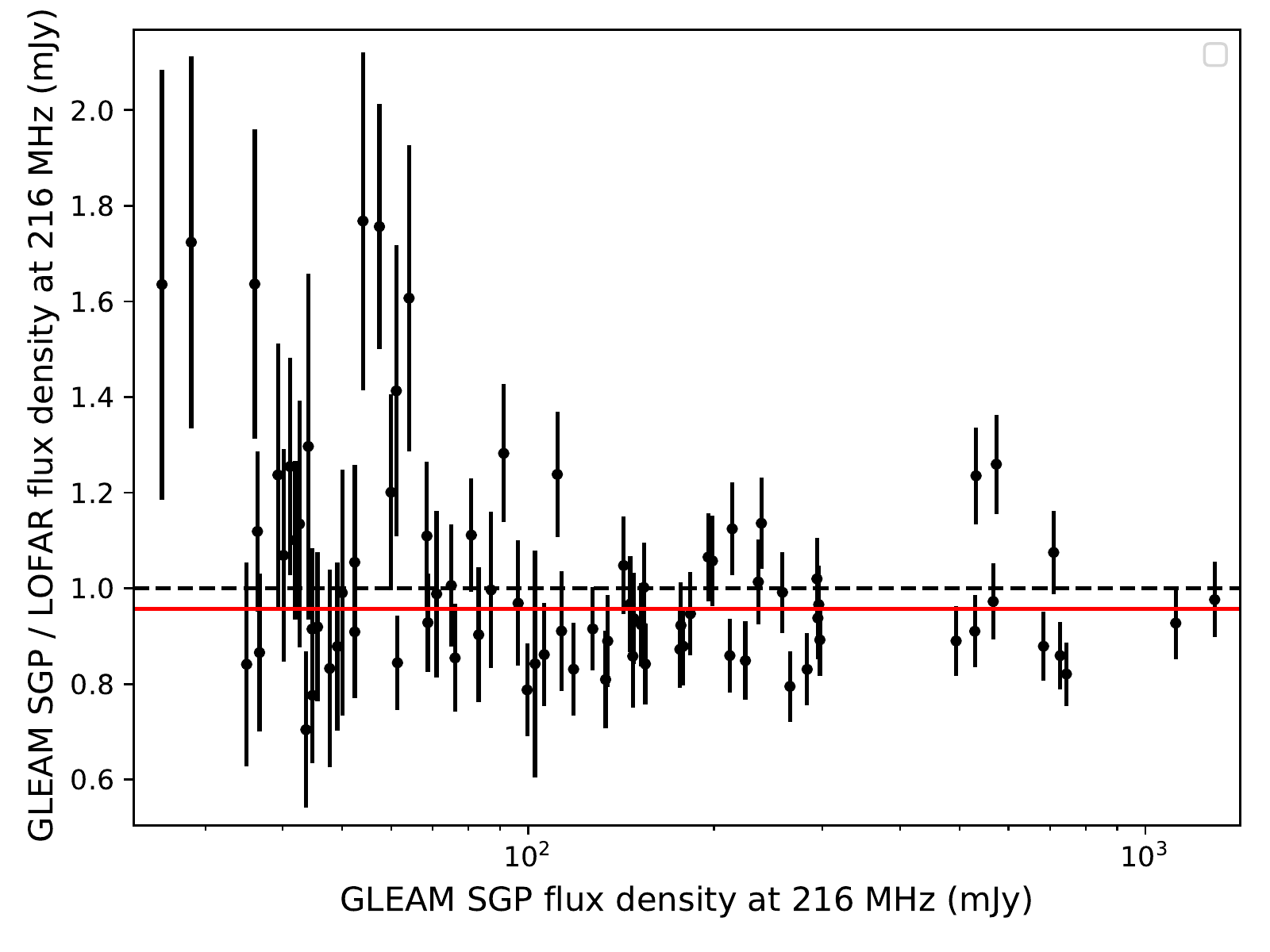}
\caption{Ratio of the GLEAM SGP to LOFAR flux density as a function of the LOFAR flux density for a sample of 82 sources in the XMM-LSS field. The GLEAM SGP flux densities are measured from the wide-band image at 216~MHz. The LOFAR flux densities originate from \citet{hale2019}. The dashed horizontal line indicates equal flux density values. The red horizontal line marks the median flux density ratio.}
\label{fig:flux_comparison}
\end{center}
\end{figure}

In Fig.~\ref{fig:flux_comparison}, the ratio, $R$, of the GLEAM SGP to LOFAR flux density is plotted as a function of the GLEAM SGP flux density. The GLEAM SGP flux densities are on average consistent with the LOFAR flux densities at the $\approx 5$ per cent level: the median value of $R$ is $0.96 \pm 0.03$ and the mean value is $1.03 \pm 0.03$. The relatively large scatter in $R$ (the standard deviation of $R$ is 0.23) is likely due to the large difference in sensitivity and resolution of the two catalogues, as well as the inclusion of sources with SNRs as low as $\approx 5$ in the GLEAM SGP image. The sources with the highest values of $R$ ($> 1.5$) all have low SNRs in GLEAM SGP ranging between 6.0 and 10.8. Their GLEAM SGP flux densities may be biased high due to the Eddington bias \citep{eddington1913} close to the survey detection limit. Another potential cause of the discrepancy is missing extended flux in the LOFAR image.

\subsection{Final catalogue}\label{Final catalogue}

The GLEAM SGP catalogue gives the position of each source in the wide-band image, and the integrated flux density and shape of each source in the wide-band and sub-band images. The local PSF at the location of each source is provided at each frequency. The catalogue contains 108,851 rows and 313 columns. Columns 1--310 are exactly the same as those in the GLEAM Exgal catalogue except that the measurements derived from the wide-band image (columns with the subscript `wide') are centred at 216~MHz rather than at 200~MHz. The remaining columns are defined as follows:

\noindent \textit{Columns 311 and 312} -- Best estimate of the 200~MHz integrated flux density, $S_{200}$, and its associated error, $\sigma_{S_{200}}$, in Jy. $S_{200}$ and $\sigma_{S_{200}}$ are set to the fitted 200~MHz flux density and its error, providing the spectrum is well fit by a power law (reduced $\chi^2 < 1.93$). If the spectrum is not well fit by a power law or the fitted 200~MHz flux density is not measured due to the low SNR, $S_{200}$ is estimated by extrapolating the 216~MHz integrated flux density from the wide-band image assuming $\alpha = -0.8$;  $\sigma_{S_{200}}$ is set to the scaled uncertainty on the wide-band flux density. These two columns are provided in order to ensure that all sources in the GLEAM SGP catalogue have a measurement of their 200~MHz flux density, as is the case in the GLEAM Exgal catalogue.

\noindent \textit{Column 313} -- Extended flag: point-like (0) or extended (1) (see Section~\ref{Extended sources}).

The electronic version of the GLEAM SGP catalogue is available from VizieR.


\section{Summary}\label{Summary}

This work presents images and an extragalactic source catalogue from combining both years of GLEAM observations at 72--231~MHz conducted with Phase I of the MWA. The data release covers a 5,113$~\mathrm{deg}^{2}$ area of sky centred on the SGP at $20\h 40\m < \mathrm{RA} < 05\h 04\m$ and $-48\deg < \mathrm{Dec} < -2\deg$. The typical rms noise level is $\approx 5$~mJy/beam and the angular resolution $\approx 2$~arcmin. The rms noise in this region of sky is $\approx 40$ per cent lower than in GLEAM Exgal, which is solely based on the first year of GLEAM observations, as a result of the longer integration times, better $(u,v)$ coverage and improved processing. A total of 108,851 source components are detected above $5\sigma$ at 216~MHz and the source catalogue includes 72--231~MHz spectral indices for 77 per cent of the components. The catalogue is estimated to have a completeness of 50 per cent at 25~mJy and a reliability of 99.99 per cent.

The GLEAM SGP data reduction procedure largely follows that of GLEAM Exgal but we make significant improvements in a number of areas: we use the GLEAM Exgal catalogue as a sky model to calibrate the snapshot data. The burden on computational resources in terms of storage space and processing time is greatly reduced thanks to the use of Dysco compression and the implementation of the Clark CLEAN algorithm in \textsc{wsclean}. We use a new full embedded element primary beam model \citep{sokolowski2017} in the calibration and mosaicking to improve the accuracy of the flux scale across the mosaics.

The GLEAM SGP data are well suited to large-scale studies of extragalactic source populations. \cite{ross2021} have searched for variable sources, including blazars and compact-steep-spectrum (CSS) sources, by comparing the flux densities and spectral shapes of sources in the GLEAM SGP year 1 and 2 mosaics. Their study represents the largest survey of low-frequency spectral variability to date, using quasi-simultaneous flux density measurements over a large fractional bandwidth. Franzen et al. (in preparation) used the GLEAM SGP data in combination with optical spectroscopy from the 6dF Galaxy Survey \citep[6dFGS;][]{jones2009} to determine the local radio luminosity function for AGN and star-forming galaxies at 200~MHz, and characterised the typical radio spectra of these two populations.

In 2018, the MWA was upgraded with the addition of a further 128 tiles, 56 of which were deployed on long baselines, doubling the maximum baseline of the array \citep{wayth2018}. In the MWA Phase II configuration, the angular resolution at 154~MHz is $\approx 1.2$~arcmin. The improvement in the angular resolution reduces the classical and sidelobe confusion limits, allowing for deeper imaging. The improvements to the data processing in GLEAM SGP are being adopted for MWA Phase II processing.

The images and source catalogue from the GLEAM SGP data release are publicly available at \url{https://data-portal.hpc.swin.edu.au/dataset/gleam-sgp-data-release}.


\begin{acknowledgements}
This scientific work makes use of the Murchison Radio-astronomy Observatory, operated by CSIRO. We acknowledge the Wajarri Yamatji people as the traditional owners of the Observatory site. Support for the operation of the MWA is provided by the Australian Government (NCRIS), under a contract to Curtin University administered by Astronomy Australia Limited. We thank the anonymous referee for helpful comments, which have substantially improved this paper. We acknowledge the Pawsey Supercomputing Centre which is supported by the Western Australian and Australian Governments.
\end{acknowledgements}

\section{Conflicts of Interest}
None.

\bibliographystyle{pasa-mnras}
\bibliography{myreferences}
\setlength{\labelwidth}{0pt}

\label{lastpage}
\end{document}